\documentclass{epl}
\usepackage{amsmath}

\def\R{{\bf R}}
\def\Itens{\mbox{\sffamily\bfseries I}}
\title{Melting-freezing cycles in a relatively sheared pair of crystalline monolayers}
\shorttitle{Melt-freeze cycles in relatively sheared crystalline layers} 
\author{Moumita DAS\inst{1}\thanks{E-mail: \email{moumita@physics.iisc.ernet.in}}, 
Sriram RAMASWAMY\inst{1}\thanks{E-mail: \email{sriram@physics.iisc.ernet.in}} 
\and 
G. ANANTHAKRISHNA\inst{2,1}\thanks{E-mail: \email{garani@mrc.iisc.ernet.in}}}
\shortauthor{M. Das \etal}
\institute{
  \inst{1} Centre for Condensed Matter Theory, Department of Physics,
Indian Institute of Science, Bangalore 560012, India\\
  \inst{2} Materials Research Centre, IISc, Bangalore 560012, India. 
}
 
\begin{document}
 
\maketitle
 
\begin{abstract}
The nonequilibrium dynamical behaviour that arises when two 
ordered two-dimensional monolayers of particles are sheared over each other 
is studied in Brownian dynamics simulations. 
A curious sequence of nonequilibrium states is observed as the driving rate 
is increased, the most striking of which is a sliding 
state with irregular alternation between disordered and ordered states.
We comment on possible mechanisms underlying these 
cycles, and experiments that could observe them.  
\end{abstract}

\begin{figure}
\twofigures[height=4cm]{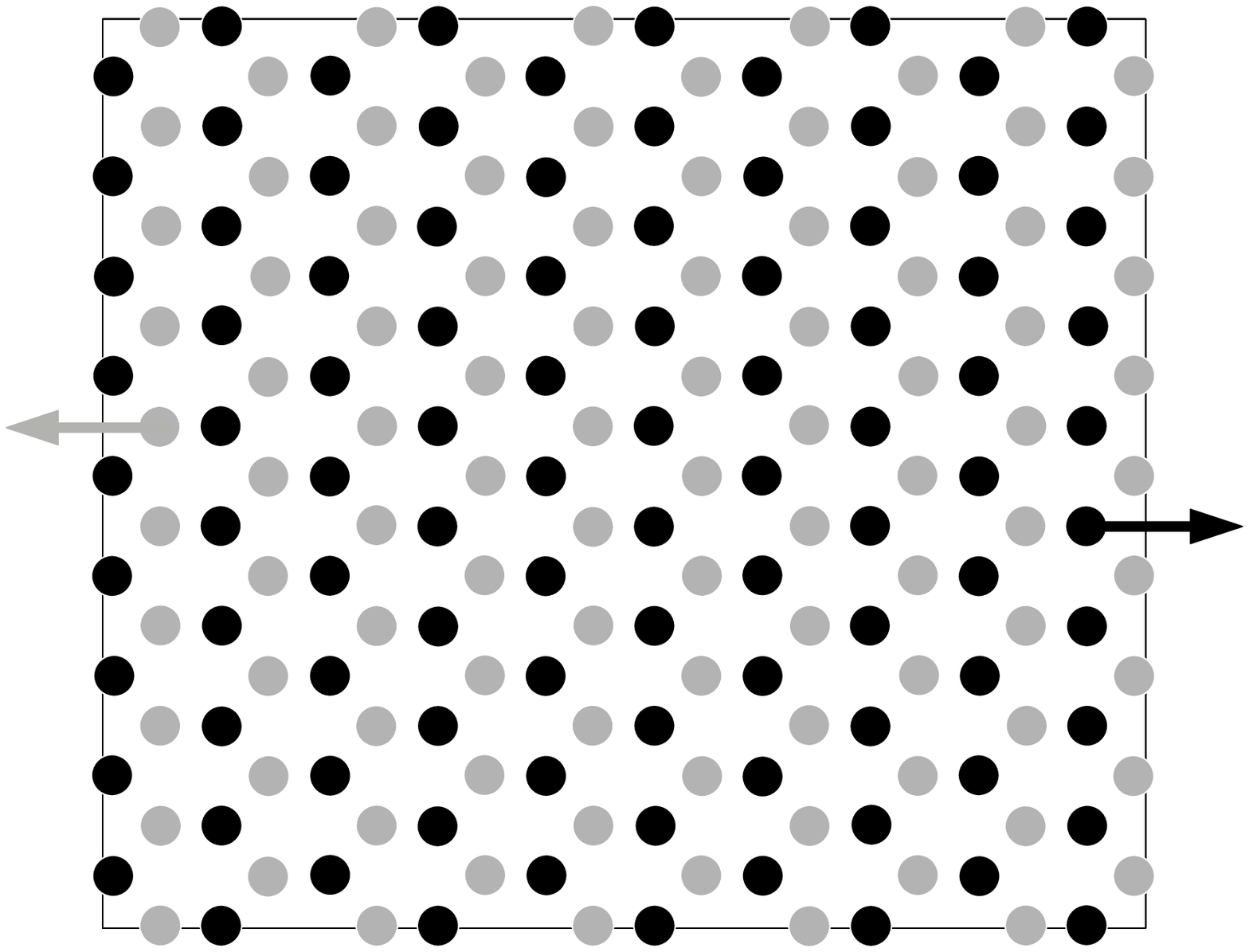}{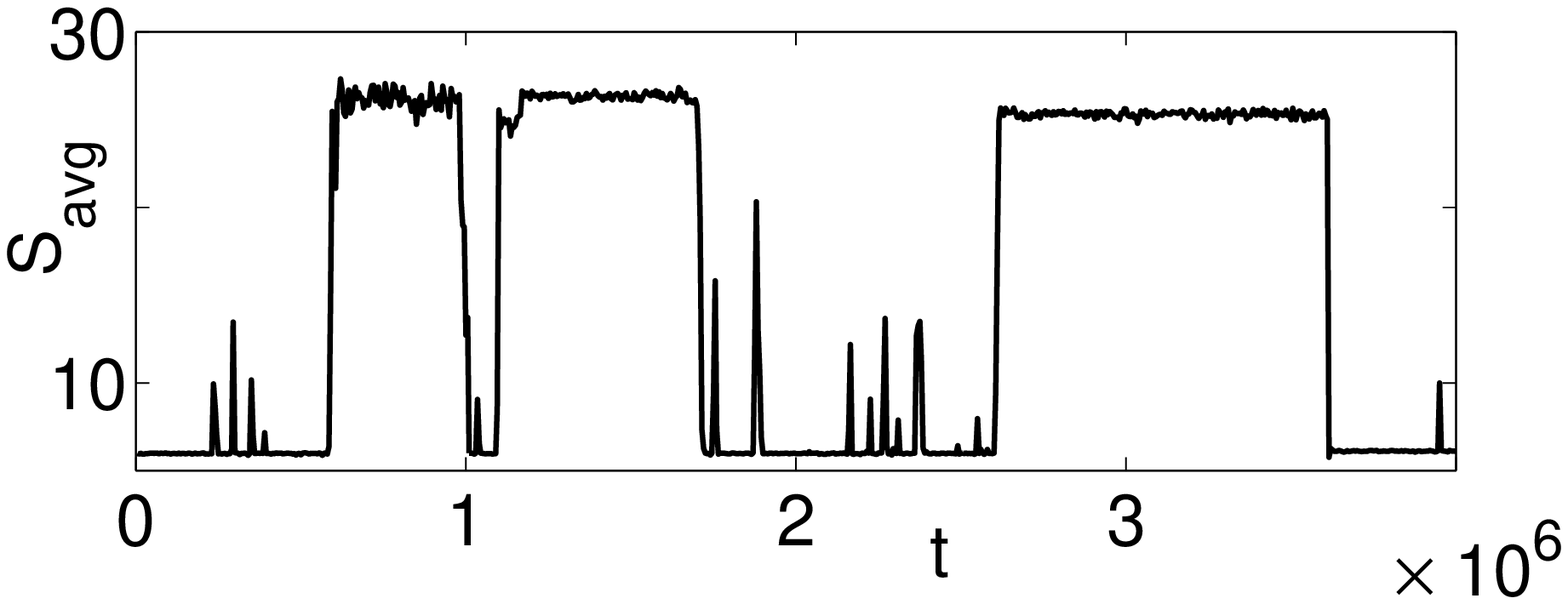}
\caption{Schematic diagram of the model.}
\label{f.model}
\caption{The structure factor height (averaged over 1st ring of maxima)
as a function of time in the melt-freeze cycle state, for $\epsilon=0.05$.}
\label{f.Sk}
\end{figure}

Solid friction \cite{b.prbk,b.mrbk,b.pr} and related problems involve the flow of ordered, 
deformable structures over or through an 
inhomogeneous medium \cite{b.balentsetal} 
which may itself be either ordered or disordered. These physical 
systems are frequently modelled theoretically in one of two ways: 
(i) a driven Frenkel-Kontorova (FK) model \cite{b.prbk,b.gr}, 
where a density wave, in the form of a ball-and-spring array, is driven by a constant 
force over a substrate modelled by a periodic pinning potential, or (ii) 
molecular \cite{b.thompsonetal} 
or Brownian \cite{b.br} 
dynamics studies of particles interacting with each other and the substrate via pair 
potentials. Among the most interesting phenomena seen in these systems are stick-slip 
\cite{b.homola,b.thompsonetal} and shear-induced melting 
\cite{b.ackerson,b.chaikin,b.sri,b.mark,b.peter}. 
It is clearly of great interest 
to find the minimal model system capable of showing phenomena such as these.
In particular: (a) does stick-slip involve crucially the influence of the 
periodic potential of the crystalline confining walls on the film of confined 
fluid ? (b) Is inertia essential? (c) Is a three-dimensional 
model necessary? In addition, the dynamics of ordered monolayers of surfactant or 
copolymer
adsorbed onto two solid surfaces sheared past each other \cite{b.copoly1,b.copoly2} 
cannot adequately be described by approaches so far discussed in the literature. 
Accordingly, we
report in this Letter a Brownian dynamics study of a model with two species of 
particles, 1 and 2, moving in two dimensions. The 11 and 22 interactions are identical, while the 12 interaction has the same form but is smaller by a 
factor $\epsilon$.
The 1 and 2 particles are driven in the $+x$ and $-x$ directions respectively under the 
action of a constant force of magnitude $F$, as in Fig.\ref{f.model}.
The physics of the third direction enters phenomenologically  
via $\epsilon$, the relative strength of the 12 interaction: as the two planes of 
particles are pushed together with increasing normal pressure, their in-plane repulsion 
and hence $\epsilon$ will increase. While preparing this paper for publication we noted 
Ref. \cite{b.lowen}, 
which studies lane formation in inter-driven interacting Brownian particles. That work 
uses a model similar to ours, but well away from the regime where either species 
can crystallise. 

\begin{figure}
\twofigures[height=4.5cm]{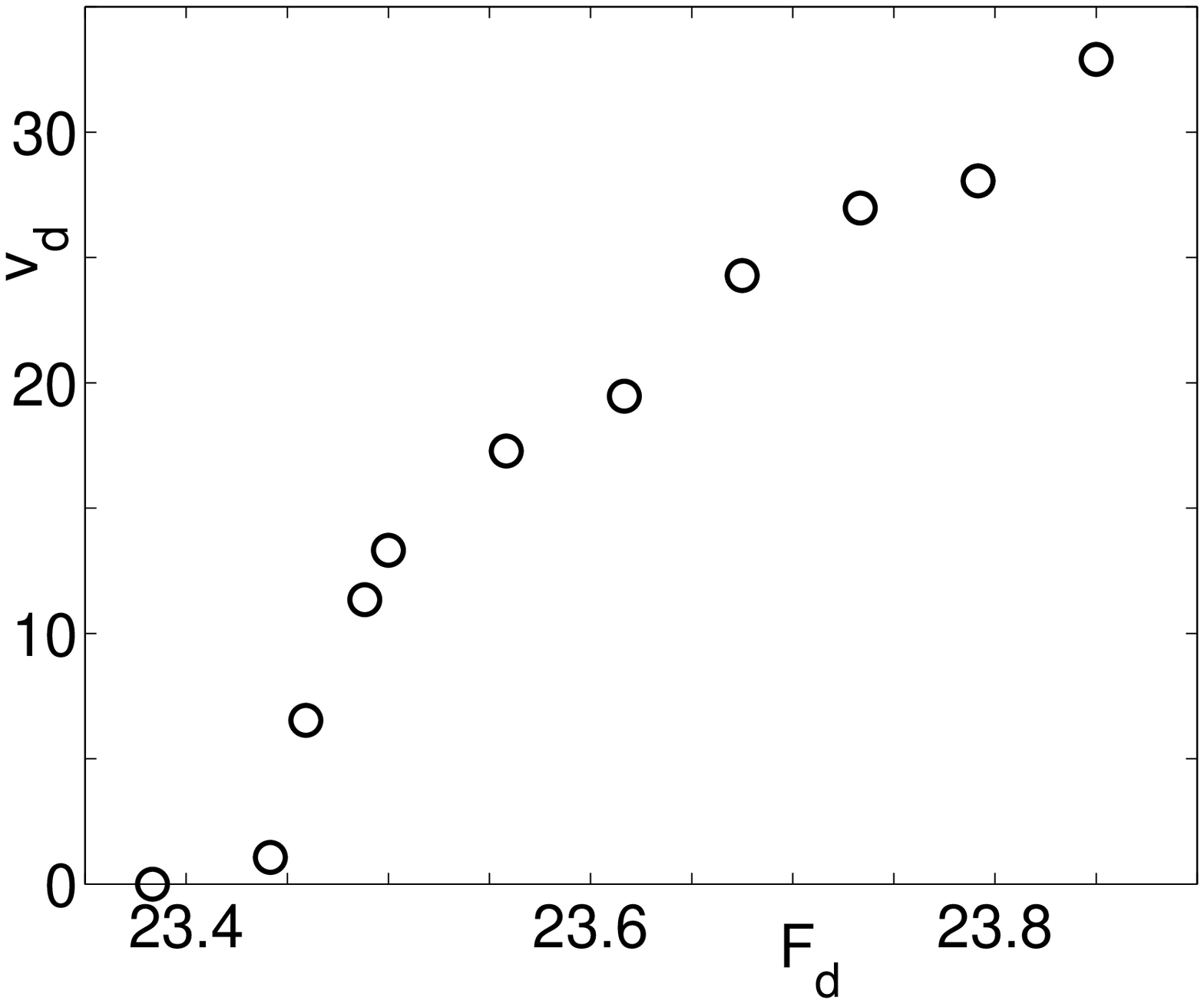}{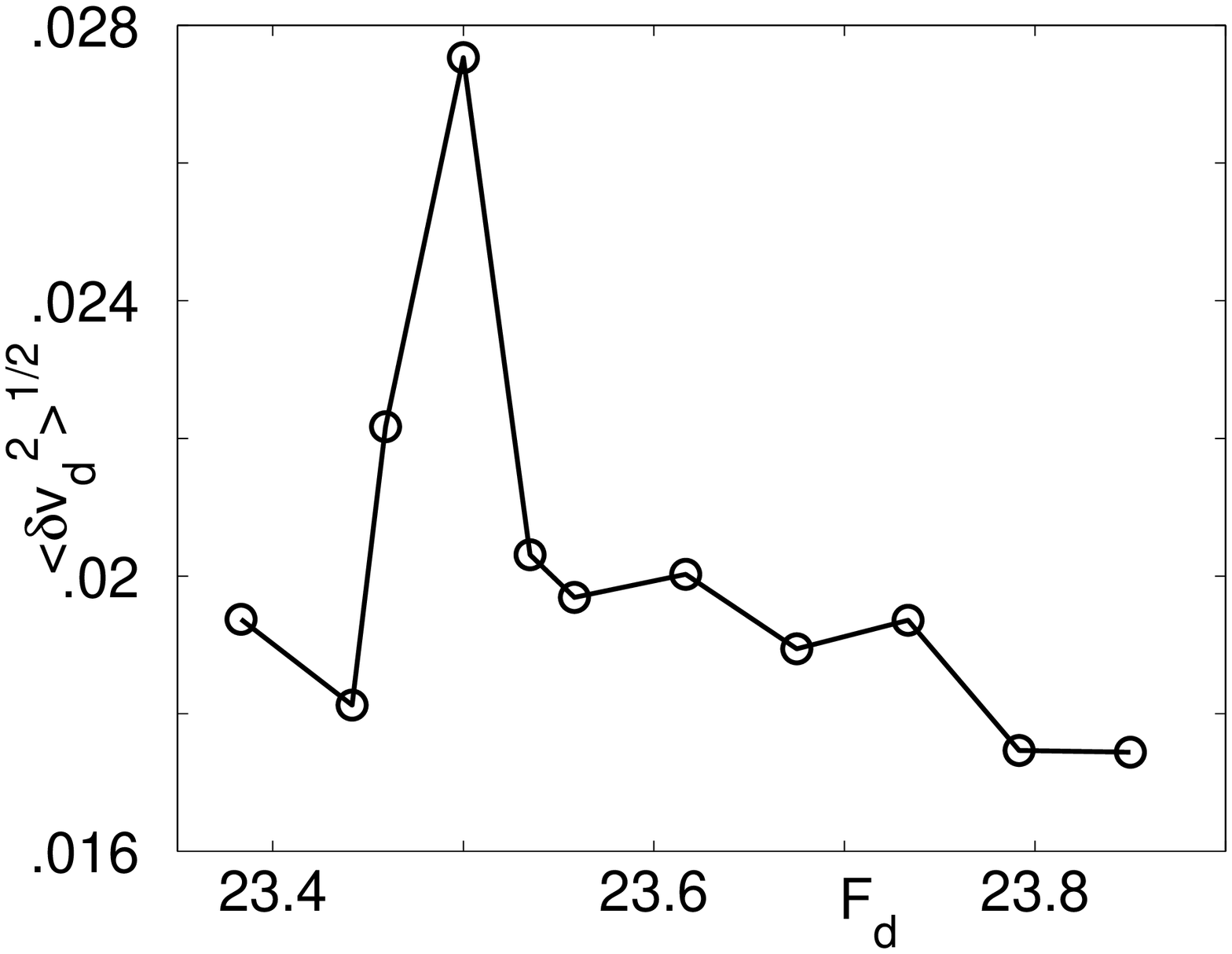}
\caption{ Mean drift velocity at onset of smooth sliding regime.}
\label{f.driftvel}
\caption{ RMS velocity fluctuations at onset of smooth sliding regime; 
$\epsilon = 0.02$.}
\label{f.RMSvelfluc}
\end{figure}

Our main results are as follows: keeping interaction strengths and temperature 
fixed, the driving force 
$F$ displays three threshold values $F_i$, $i = 1, 2,3$. Our most striking 
observations are the melt-freeze cycles (Fig.\ref{f.Sk}),  
to which we shall return. In sequence, however: For $F < F_1$, 
the drift speed $v_d$ is effectively pinned at zero. In this regime, an initial poorly 
ordered configuration displays transient motion 
while settling down into a macroscopically ordered state, and then ceases to drift. For all 
$F > F_1$, $v_d > 0$, 
with a smooth onset (Fig.\ref{f.driftvel}) and enhanced velocity fluctuations 
(Fig.\ref{f.RMSvelfluc}) 
at $F_1$. For $F_1 < F < F_2$ as well as for $F > F_3$, both 1 and 2 components are 
well-ordered, sliding crystals. For $F_2 < F < F_3$ we find striking cycles of melting 
and freezing, as signalled by the alternating growth and decay of 
the peaks in the structure factors or pair correlation functions for either species 
(Fig.\ref{f.Sk}). In the course of these irregular cycles the observed structures 
range from highly crystalline to (anisotropic) liquid-like or perhaps 
smectic-like, and onsets of the change 
between these two states are not necessarily simultaneous for 1 and 2. Unlike in 
\cite{b.thompsonetal}, it can be seen from the figure that the system 
spends {\em comparable} amounts of time in the ordered 
and disordered states. 
With time, columns of particles aligned normal to the mean drift start to undulate 
and, when this undulation builds up sufficiently, the whole system disorders abruptly.  
The thresholds $F_i$ 
depend, of course, on the pair interactions.
We are unaware of any other driven systems with {\it neither inertia nor 
a lubricating fluid film} which display such alternating {\it cycles of order 
and disorder as the persistent long-time state}. 
We defer to the end of 
this paper our qualitative explanation of the melt-freeze cycles, as well as an outline 
of possible experimental tests. Let us now present our study and its results in 
more detail. 

The positions $\{\R^i(t)\}$, where the superscript $i = 1, ... N$ labels the particle, 
evolve according to overdamped Langevin equations, with independent Gaussian, zero-mean, 
thermal white noise sources ${\bf h}^i$, 
interparticle forces ${\bf f}^i$ from the pair potentials, and 
equal and opposite constant external forces 
$\pm F \hat{\bf x}$ on the 1 and 2 species respectively. 
Let us nondimensionalise our variables as follows: 
scale all lengths by $\ell = (2\sqrt{3}n_0)^{-1/2}$, 
where $n_0$ is the mean number density of either species,
 energy by Boltzmann's constant $k_B$ times 
temperature $T$ (and hence force by $k_BT/\ell$), 
and time by $\tau \equiv \ell^2/D$, where $D$ is the Brownian diffusivity.   
Then our nondimensional discretised Langevin equations are   
\begin{eqnarray}
\label{e.langnondim}
{\R}^i(t + \delta t) = {\R}^i(t) + \delta t [{\bf F}^i +  {\bf f}^i({\R}(t)) + {\bf h}^i(t)]
\end{eqnarray}
where ${\bf F}^i = (-1)^{\alpha_i}F \hat{\bf x}$ 
is the external driving force on 
the $i$th particle of type $\alpha_i$ (= 1 or 2), 
${\bf f}^i(\R^i) = - \sum_{j \neq i} 
\nabla V_{\alpha_i \alpha_j}(\R^i - \R^j)$,  
and $\langle {\bf h}^i(0) {\bf h}^j(t) \rangle = 2 \Itens \delta^{ij} \delta(t)$, 
$\Itens$ being the unit tensor\footnote{The dynamics will in general generate 
{\em nonthermal} noise   
which will emerge upon coarse-graining, and does not need to be included 
by hand in our microscopic description. This 
renormalised noise will have anisotropic 
correlations with a strength determined by a combination of the driving force 
and $k_BT$.}.   
The dimensionless pair potentials have the screened Coulomb form 
$V_{11}(r) = V_{22}(r) = \epsilon^{-1}V_{12}(r) 
= (U/r) \exp(-\kappa r)$ at interparticle separation ${\bf r} = (x,y)$, 
where the subscripts indicate which types of particles are interacting. 
The dimensions $L = (\sqrt{3}/2)\times 20$ and $W = 20 $ of 
our rectangular box are such   
that for $\epsilon = 0$ a triangular lattice of each species fits in the box, 
and $U =1.75 \times 10^4$ and $\kappa \ell = 0.5$ are chosen so that the equilibrium phase for $\epsilon =0$ is such a 
triangular lattice\footnote{If the two species differed in such a way as to form 
lattices which were mutually incommensurate, very different physics could be expected, 
which we do not discuss here.}.  
We first discuss results for a system with $N= 200$ particles, i.e., $100$ 
of each species, with $\epsilon = 0.02$. Later in the paper we shall mention 
observations for other system sizes
and values of interspecies interaction wherever relevant.
 
The dimensionless time-step $\delta t = 6.5 \times 10^{-6}$, and the results we report 
are mainly for runs of $10^6$ such steps. Over this time, the 1 and 2 lattices 
sweep through each other a few to several hundred times depending upon the
drive. 
In order to drift under the action of the driving force $F$ the particles have to 
overcome a barrier of order $V_{12}(\ell)$ over a distance of order $\ell$. 
Thus, although $F$ is itself dimensionless, we state our results in terms of the 
physically relevant dimensionless combination 
$F_d \equiv F \ell /V_{12}(\ell)$.  
We monitor the structure and dynamics of the system through snapshots of configurations, 
drift velocities $v_d$, particle-averaged local velocity variances 
$\langle (\delta v)^2 \rangle$, 
pair correlation functions $g_{\alpha \beta}({\bf r})$ as functions of separation 
${\bf r}$, and static structure factors $S_{\alpha \beta}({\bf q})$ as functions 
of wavevector ${\bf q}$, where $\alpha$ and $\beta$ range over 1, 2. Let us 
discuss the ``phases'' seen as $F_d$ is increased, keeping other parameters 
fixed at values $\epsilon=0.02$ and $\kappa \ell=0.5$. The typical 
initial state of the system at $F_d=0$ is an imperfectly ordered crystal. The 
application of a small nonzero $F_d$, well below that required for macroscopic 
relative motion 
of species 1 and 2, is seen to produce movement in regions where particles were 
initially in unfavourable positions. After these transient motions the system settles 
down into a highly annealed structure with both 1 and 2 components showing 
near-perfect long-range crystalline order, with no further relative drift 
of 1 and 2 except presumably an activated creep which we cannot detect. This ``phase'' 
does not seem to 
have any striking properties so we shall not discuss it further. The first 
nonequilibrium steady state of interest is seen when $F_d$ crosses the first threshold 
$F_1$, whereupon the 1 and 2 components acquire a macroscopic 
relative drift velocity $v_d$ as shown in Fig.\ref{f.driftvel}.  
The two components slide smoothly past each other in lanes of width equal
 to the interparticle distance, with negligible distortion or disorder. 
A large but finite enhancement in $\langle (\delta v)^2 \rangle$  
is clearly seen (Fig.\ref{f.RMSvelfluc}) at the onset of this smooth sliding state,  
indicating large-scale inhomogeneous depinning. 
This does not tell us in detail about the character of the depinning transition since 
$\langle (\delta v)^2 \rangle$ can be finite even if the small-wavevector velocity 
variance (which we have not measured) diverges. 
Detailed measurements of the spatial and temporal correlations of the 
velocity are required 
before we can say more about the nature of this depinning \footnote{which 
should be a strong crossover rather than a true transition.}\cite{b.middleton}.  

\begin{figure}
\onefigure[height=3.1cm]{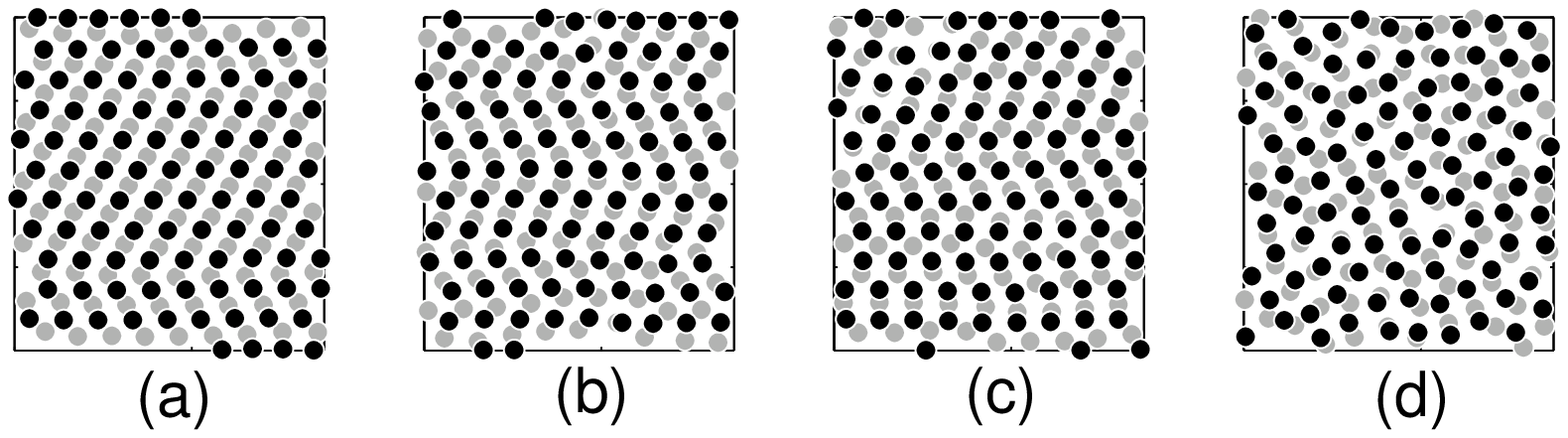}
\caption{ From order with kinks (a) to disorder (d) over a short time in the
Melt-Freeze cycle. Black and grey particles move to the right and the left 
respectively; $\epsilon = 0.02$, $F_d = 40.83$. }
\label{f.stripemelt}
\end{figure}

\begin{figure}
\leavevmode
\twofigures[width=6cm]{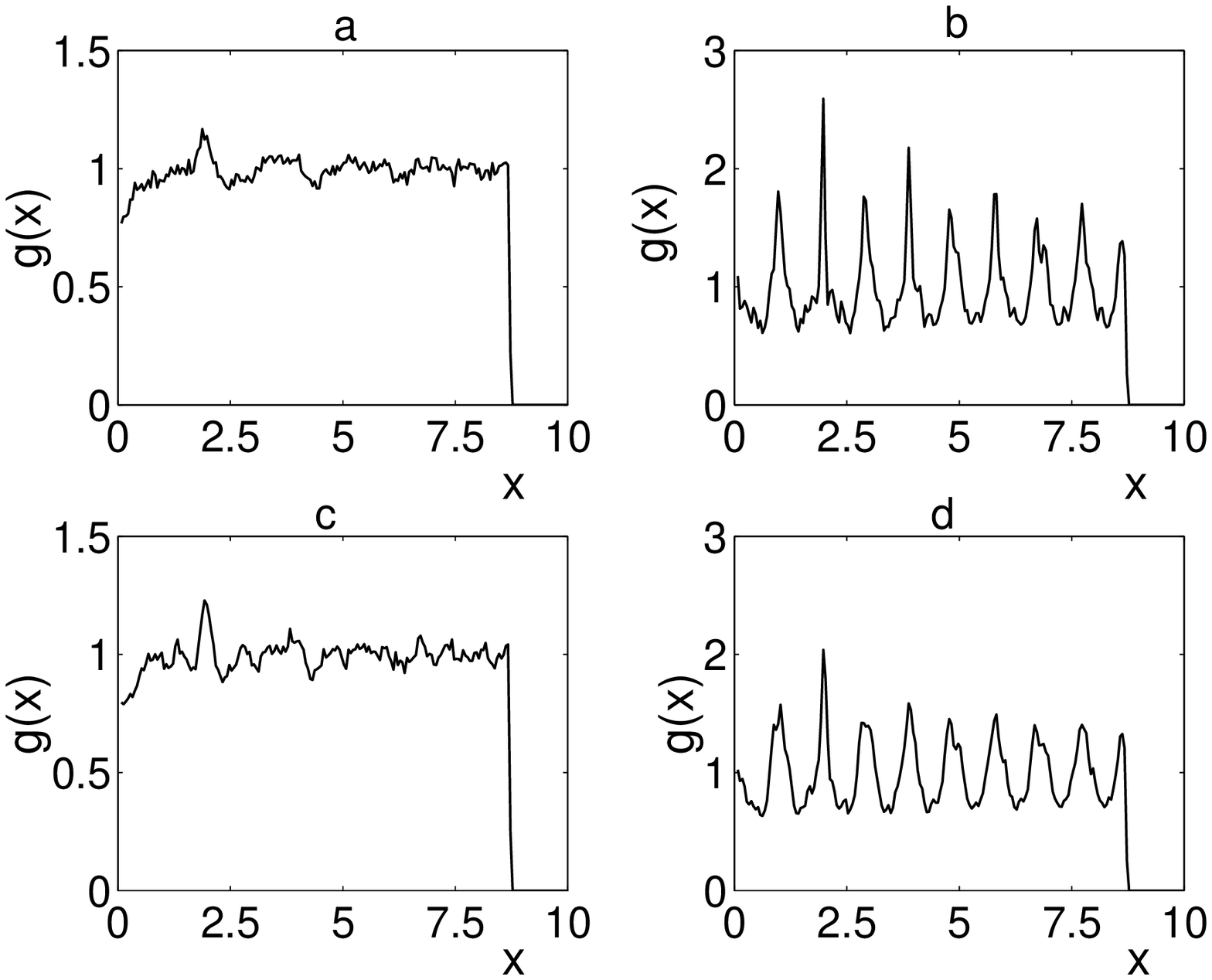}{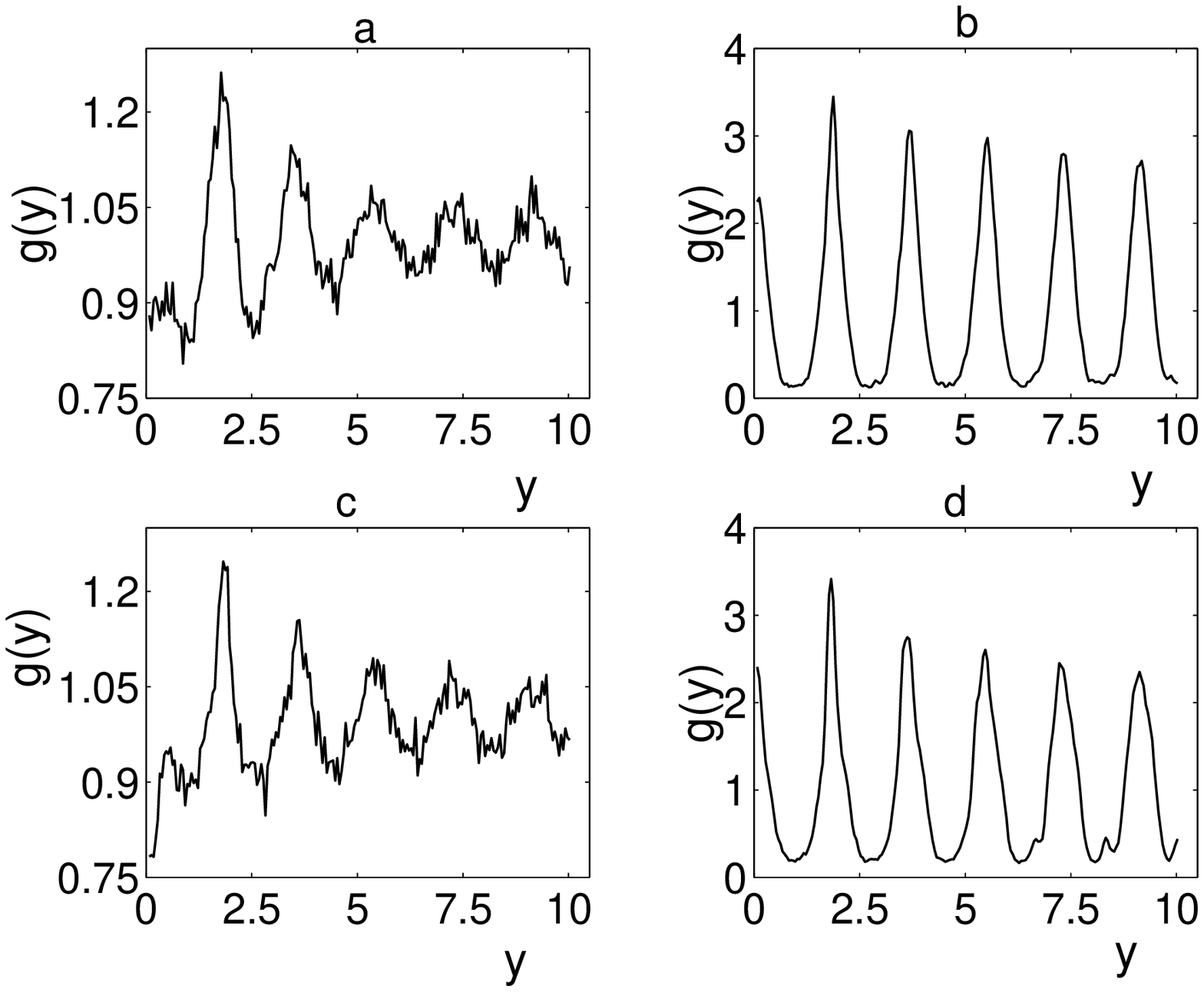}
\caption{Time evolution of $g(x)$ for species 1. Each figure is an average 
over 100 configurations evenly spaced in time over a span of 5000 time steps, 
for $\epsilon = 0.02$, $F_d =40.83$.  
The final times for (a), (b), (c) and (d) correspond respectively to 
$9.5\times 10^4$, $8.65\times 10^5$, $1.01\times 10^6$ and $1.39\times 10^6$ 
timesteps.}
\label{f.gxaa}
\caption{Time evolution of $g(y)$ for species 1; times, parameter values 
and averaging as in Fig. 
\ref{f.gxaa}.} 
\label{f.gyaa}
\end{figure}

Upon further increasing $F_d$ the melt-freeze cycles mentioned above appear. 
These cycles are our most important observation, so let us present their 
features in some detail. They are seen most strikingly in the time-dependence of the 
peak height of the (short-time averaged) static structure factor $S({\bf k})$ as shown 
in Fig.\ref{f.Sk}, which alternates between long stretches of crystal-like and 
comparably long stretches of liquid-like values as the simulation progresses. 
This is in contrast with the behaviour seen in \cite{b.thompsonetal}, 
where the time spent in the disordered state is much smaller, as though the system 
preferred order to disorder.  
The structure factor height in the ordered part of the cycle is proportional to 
the number of particles, so that we are justified in terming this regime 
crystalline. 
The cycle persists without limit in time, so far as we can tell. 
For $\epsilon = 0.02$ the threshold $F_d = F_2 =40.54 $. 
For slightly {\em higher}  values of $\epsilon$, say 0.03, 0.04, 0.05 
we find the the stretches of 
ordered and disordered behaviour are better defined than for 0.02. 
In addition, increasing $\epsilon$ slows down the onset of order 
during the cycle. The drift velocity correspondingly 
is large in the disordered and small in the ordered part of the cycle. 
As can be  inferred from Fig.\ref{f.Sk}, this is not at all like 
the sawtooth stick-slip seen, e.g., in \cite{b.thompsonetal}. 
Both velocities and order parameters grow and decline abruptly.   
There is a curious metastability 
associated with the cycles: if the initial state is 
chosen to be a {\em perfectly ordered lattice} as in Fig.\ref{f.model}, 
a disordered configuration fails to 
nucleate over the largest time we are able to simulate. 
Both the cycles and the ordered sliding states thus seem to be locally stable. 
But if we disturb this initial perfectly ordered lattice by moving a single particle by, 
say, one lattice spacing, the melt-freeze cycles resume.  
In the ordered part of the cycle the smooth relative motion of the 1 and 2 lattices 
is disturbed now and then by kinks, i.e., a row moving out of step with 
adjacent rows, as seen in Fig.\ref{f.stripemelt}a. 
The positional variance along $x$ increases (Figs.\ref{f.stripemelt}b and c) 
and, when it gets large enough, the system
abruptly transits to a disordered state (Fig.\ref{f.stripemelt}d). 
This state persists for a long time, before order once again 
sets in.  The cycles are seen clearly 
in the time evolution of the pair distribution functions for 
separations along and transverse to the direction of the drive 
(Figs.\ref{f.gxaa} and \ref{f.gyaa}).
Note that the order along $y$ is never totally destroyed; larger simulations 
are required to establish its character over the course of the cycles. 
Fig.\ref{f.stripemelt} suggests that the disordering mechanism can also be regarded as 
excess fluctuations, with wavevector along $y$, in the phase of the 
density-wave component with wavevector along $x$.  
The cycles are asymmetrical: the order grows appreciably faster than it decays
but both growth and decay are over timescales much shorter than the ordered
or disordered stretches. The melt-freeze cycles for the two species are not necessarily simultaneous.
Typically one species begins to order while the other is still disordered. 

Let us consider briefly the the effect of finite size.  
Inspection of Fig.\ref{f.stripemelt}d (an image from the disordered 
part of the cycle) shows strong residual correlations, most easily seen by 
looking along a grazing angle. This 
is consistent as well with the form of $g(y)$ in Fig.\ref{f.gyaa}a and c.  
This indicates that the correlation length does not drop below about 4 interparticle 
spacings, so that the difference between ordered and disordered states will 
be hard to detect for small system sizes. Indeed, we are unable to see 
`cycles' for  $N \leq 64\times2$.  
Simulations of larger systems, $N = 144\times2, \, 169\times2$ and $256\times2$ 
show the same qualitative behaviour as for $N = 100\times2$ particles,
with less than a 1 percent change in the range of values of $F$ for 
which the ``cycles'' are seen. 

\begin{figure}
\onefigure[height=4.4cm]{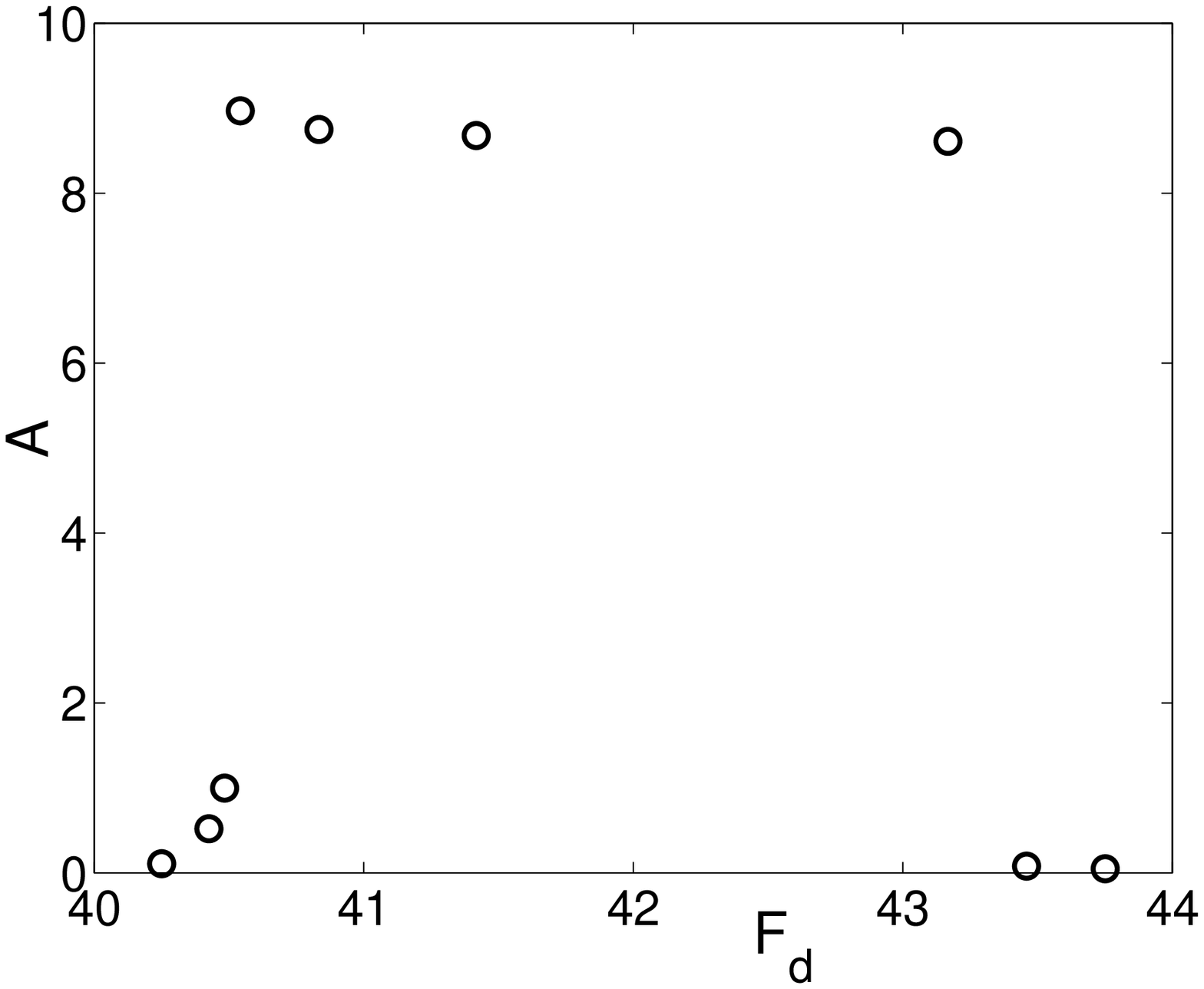}
\caption{Amplitude of the melting freezing cycle  
as a function of driving force $F_d$, for $\epsilon=0.02$.}
\label{f.amplitude}
\end{figure}

We also observe that an increase in 
the relative strength $\epsilon$ of the $1-2$ interaction  
slows down the onset of order substantially. For
$\epsilon = 0.05$, starting from a random initial configuration, 
we observed only a few cycles over the simulation
 run, and none for $\epsilon = 0.10$.
Fig.\ref{f.amplitude} shows the amplitude $A$ of the cycle (an ``order parameter'' 
for this behaviour), defined as the difference in the value of S in the 'ordered'
and 'disordered' regimes, averaged over the duration of the cycle, as a function of 
$F_d$ for fixed values of the other parameters. 
 
Remarkably, the cycles are seen only in a limited 
range of $F_d$, disappearing again when $F_d$ crosses another threshold $F_3$.  
Beyond $F_3$, the structure again consists of ordered relatively sliding 
1 and 2
lattices. This phase appears to be of the same nature as that below 
$F_d = F_2$. If we lower $F_d$ below $F_2$ after the system has settled into a 
steady state, we find, as we remarked above, that the melt-freeze cycles do not appear 
unless at least one particle is displaced sufficiently from its ideal ordered location. 
Thus, the transition into the melt-freeze cycle phase is characterised by metastability 
and hysteresis.  
   
We note here that the re-appearance of a smooth sliding state resembles 
the re-entrant ordered state seen in \cite{b.mark,b.sri}. Possibly,  
therefore, the melting-freezing cycles are the analogue, in our system, 
of the phenomenon of shear-induced melting 
\cite{b.ackerson,b.chaikin,b.mark,b.sri,b.peter}. 

Having established the reality of the melt-freeze cycles 
as a distinct, reproducible type of behaviour in relatively driven lattices, 
let us try to explain why they occur. When the two arrays of particles are 
pushed through each other, there is a competition between two timescales. 
The arrays traverse one lattice spacing $a$ in a time $\tau_1$ given by $a/v_d$. 
The slowest local relaxation will be that of species-1 particles in the potential 
well provided by their species-2 neighbours (and vice versa). The timescale $\tau_2$ 
for this 
will be of order the friction coefficient divided by the curvature of the local 
1-2 potential. 
If $\tau_1 >> \tau_2$, the lattices of the 
two species have 
plenty of time to relax as they interpenetrate, so that we find smooth, orderly 
sliding. If instead $\tau_1 << \tau_2$, each species averages over the periodic 
potential of the lattice of the other species, so that again we expect smooth 
sliding. The maximum mutual disruption of the two lattices is expected only 
in a window of parameter values where the two times are comparable. When 
this condition is met, a species-2 unit cell, say, is distorted by successive 
passages of species-1 particles, resulting in total disruption of the density-wave 
components with wavevector along $x$. The resulting structure probably retains 
some order along $y$ (see Fig.\ref{f.gyaa}), so that each species perhaps provides 
a weak periodic potential along $y$ for the other species. Through a mechanism similar 
to that in \cite{b.chowdhury}, this can once again induce order along $x$ as well.  
Once this happens the movement of the lattices will again cause disruptions, and 
the cycle then continues. 
For our parameter values, $\tau_1 \simeq 0.00022$ in the regime $F_2 < F < F_3$, and  
$\tau_2 \simeq 0.00019$ for $\epsilon= 0.05$, which suggests the two scales should 
be in competition in the melt-freeze regime, consistent with our arguments above.  
This reasoning 
is of course simplistic, since the ordered or disordered nature of a state is 
determined only in the limit of infinite size, where long-wavelength collective 
modes will play an important role. It is nonetheless reassuring that our crude estimates 
above are consistent with our observations.    

Let us take these arguments further:
In the presence of the constant mutual driving force the system alternates 
in time between a situation where 
interpenetrating triangular lattices are favoured and one where such a state is 
hard to accommodate. This suggests a possible relation to stochastic resonance
\cite{b.stochres}, 
and indeed Fig.\ref{f.Sk} is strongly reminiscent of the time-series of a particle  
in a periodically modulated bistable potential.  
Further work is in progress to make this connection more precise. 

In conclusion, we have studied a rather special nonequilibrium binary system 
in which the two species, separately in crystalline arrays, are driven through each 
other in opposite directions. We predict 
a rich range of nonequilibrium effects, the most striking of which are irregular 
cycles of melting and freezing. Such systems can be experimentally realised, 
generalising the ideas of \cite{b.lowen}, 
by starting with a compound colloidal crystal made of two species of differently charged 
colloidal particles and applying a constant
external electric field. Measurements made in the centre-of-mass
frame would mimic two species being driven in opposite directions.
Another possibility would be to shear past each other two dissimilar 
solid surfaces patterned 
with ordered copolymer structures \cite{b.copoly2} or colloidal particles. 
The copolymer or colloids 
would have to be of two kinds, one of which adsorbed preferentially onto each of the 
two surfaces.    
We expect that similar effects should arise in sheared colloidal crystals as well, 
where adjacent crystal planes play the role of the two crystalline layers. Lastly, 
although the basic ideas and observations presented here could conceivably 
play some role in solid friction as well, the driving speeds required 
might be proportional to elastic wave speeds, making the effects inaccessible 
to experiment.  

\acknowledgments          
MD thanks CSIR, India for financial support, and C. Dasgupta for useful 
discussions, and GA acknowledges support from DST grant SP/S2/K-26/98.  
                        
\end{document}